# A simulation study comparing likelihood and non-likelihood approaches in analyzing overdispersed count data[*]

**Stanley Xu** [??]    and **Gary Grunwald** [†]

*Institute for Health Research, Kaiser Permanente Colorado, Denver, CO 80237.*

*Department of Preventive Medicine and Biometrics, School of Medicine, University of Colorado Health Sciences Center, Denver, Colorado 80262.*
*e-mail:* `stan.xu@kp.org`; `Gary.Grunwald@UCHSC.edu`

**Richard Jones** [‡]

*Department of Preventive Medicine and Biometrics, School of Medicine, University of Colorado Health Sciences Center, Denver, Colorado 80262.*
*e-mail:* `Richard.Jones@UCHSC.edu`

**Abstract:** Overdispersed count data are modelled with likelihood and non-likelihood approaches. Likelihood approaches include the Poisson mixtures with three distributions, the gamma, the lognormal, and the inverse Gaussian distributions. Non-likelihood approaches include the robust sandwich estimator and quasilikelihood. In this simulation study, overdispersed count data were simulated under the Poisson mixtures with the gamma, the lognormal and the inverse Gaussian distributions, then analyzed with the five likelihood and non-likelihood approaches. Our results indicated that 1) when the count data are mildly overdispersed, there are virtually no differences in type I error rate, standard error of the main effect, and empirical power among the five methods; 2) when the count data are very overdispersed, none of these five approaches is robust to model misspecification as evaluated by type I error rate, standard error of the main effect, and empirical power. This simulation study raises caution on using non-likelihood method for analyzing very overdispersed count data because of likely higher type I error and inappropriate power levels. Unlike non-likelihood approaches, likelihood approaches allow for statistical tests based on likelihood ratios and for checking model fit and provide basis for power and sample size calculations. When likelihood approaches are used, we suggest comparing likelihood values to select the appropriate parametric method for analyzing very overdispersed count data.

## 1. Introduction

---

[*]First supporter of the project
[†]Second supporter of the project
[‡]Third supporter of the project

Poisson regressions are widely used in analyzing count data which often arise in medical research such as counts of adverse events, number of episodes of a chronic disease, and utilization of a health care system (Neas, Schwartz, and Dockery, 1999; Thall and Vail, 1990; Estabrooks, Nelson, Xu, et al, 2005; Sills, Shetterly, Xu, et al, 2007). A complication in Poisson regressions is that the assumption of mean equal to variance may be violated. Although it is rare, underdispersion (variance less than mean) can happen (Efron, 1986; Faddy, 1997; Ridout and Besbeas, 2004). More commonly, if the variance is greater than the mean, the count data are overdispersed. Failure to deal with the dispersion results in false conclusions, often falsely rejecting the null hypothesis although the parameter estimates from maximum likelihood are consistent and efficient (Paul and Plackeet, 1978; Cox, 1983).

There are two classes of approaches to deal with overdispersion in count data analyses: non-likelihood and likelihood. Non-likelihood approaches such as the robust sandwich estimator (Huber, 1967; White, 1982) and quasilikelihood (McCullagh and Nelder, 1989) do not assume a probability distribution of the unobserved heterogeneity which causes overdispersion. Likelihood approaches use a Poisson mixture with a non-negative variable, $\nu$, which is often called the extra-variation factor. Because of its importance and wide use, a log link function between independent variables and the mean of the dependent variable is assumed in this paper. Let $\mu_j = \exp(\mathbf{x}_j\boldsymbol{\beta})$ where $\boldsymbol{\beta}$ is a column vector of the fixed regression coefficients to be estimated, and $\mathbf{x}_j$ is a row vector of covariates for the $j$th subject. $\mu_j$ is the Poisson distribution parameter without overdispersion. With overdispersion, the Poisson distribution parameter can be modeled as (Cameron and Trivedi, 1998, p100),

$$\lambda_j = \nu_j\mu_j = \nu_j \exp(\mathbf{x}_j\boldsymbol{\beta}). \tag{1}$$

Different distributions of $\nu$ with mean 1 and variance $\sigma_\nu^2$ have been proposed to take into account overdispersion when count data are analyzed, including the gamma distribution (Manton, Woodbury, and Stallard 1981; Margolin, Kaplan, and Zeiger 1981), lognormal distribution (Hinde, 1982), and inverse Gaussian distribution (Sichel, 1971, 1973 and 1974; Houggard, Lee, and Whitmore, 1997).

However, only a few studies have compared these methods based on limited evaluation measures (Breslow, 1990; Lee and Nelder, 2000). There is not a systematic comparison of all these likelihood and non-likelihood approaches based on broad evaluation measures. Theoretical comparison is not available because there are no closed form solution for coefficients for methods such as likelihood methods requiring numerical integration and non-likelihood methods. To address the gap in the current literature, we conducted a simulation study which evaluates these approaches based on type I error rate, bias, standard error of main effect, and empirical power. The likelihood and non-likelihood methods are reviewed in section 2. In section 3, count data are simulated under various assumptions and analyzed with these five approaches.

## 2. Overdispersion models

Let $y$ have a Poisson distribution with parameter $\lambda$ as in (1). Then the mean and variance of $y$ are $E(y) = \mu$ and $\text{Var}(y) = \mu + \text{Var}(\nu)\mu^2$.

### 2.1. Gamma distribution of $\nu$

Let $\nu$ in (1) have a gamma distribution with mean 1 and $\text{Var}(\nu) = \tau$,

$$g(\nu|\tau) = \frac{1}{\Gamma(\frac{1}{\tau})\tau^{\frac{1}{\tau}}}\nu^{(\frac{1}{\tau}-1)}\exp(-\frac{\nu}{\tau}).$$

The marginal distribution of $y$ is

$$h(y|\mu,\tau) = \int \frac{\exp(-\nu\mu)(\nu\mu)^y}{y!}g(\nu|\tau)d\nu = \frac{\Gamma(y+\frac{1}{\tau})(\tau\mu)^y}{y!\Gamma(\frac{1}{\tau})(1+\tau\mu)^{y+\frac{1}{\tau}}}$$

which is the negative binomial distribution. This well known result is derived by Cameron and Trivedi (1998) citing references back to 1920.

When $\tau \to \mathbf{0}$, the negative binomial distribution approaches the Poisson distribution. For an observation, $y_j$, the log likelihood $(ll_j)$ is

$$
\begin{aligned}
ll_j =\ & y_j\log(\tau\mu_j) - (y_j + 1/\tau)\log(1+\tau\mu_j) \\
& + \log(\Gamma(y_j + 1/\tau)) - \log(\Gamma(1/\tau)) - \log(\Gamma(y_j + 1)).
\end{aligned}
$$

This is a useful computing formula since most computing packages have the log of a gamma function as a standard function. Lawless (1987) examined the efficiency and robustness properties of this Poisson mixture with $\nu$ following a gamma distribution (GM method).

### 2.2. Lognormal distribution of $\nu$

Another distribution for the extra-variation factor, $\nu$, in (1) is lognormal. Few studies have been carried out using the lognormal distribution for this purpose (Hinde, 1982). However, it has been used in analyzing correlated non-Gaussian variables (Stiratelli, Laird and Ware, 1984; Pinheiro and Bates, 1995). By including a random coefficient, $\gamma_j$, and an unknown constant, $\frac{\sigma^2}{2}$, in the fixed effect as in equation (2), the potential overdispersion in count data is taken into account. $\gamma$ is assumed to be $N(0, \sigma^2)$. Then

$$\lambda_j = \exp(\mathbf{x}_j\boldsymbol{\beta} + \gamma_j - \frac{\sigma^2}{2}) = \nu_j\mu_j \tag{2}$$

where $\nu_j = \exp(\gamma_j - \frac{\sigma^2}{2})$ has a lognormal distribution with mean equal to 1 and $\text{Var}(\nu) = \exp(\sigma^2) - 1$ based on the properties of the lognormal distribution (Casella and Berger, 1990, pp110-111).

The above form of $\lambda_j$ in (2) is very similar to the negative binomial distribution except that $\nu_j = \exp(\gamma_j - \frac{\sigma^2}{2})$ is a lognormal distribution here, while $\nu_j$ is a gamma distribution in the negative binomial case. In fact, the lognormal distribution is similar in shape to the gamma distribution (Casella and Berger, 1990, p111).

The marginal probability density function of the observation has no closed form but can be approximated by numerical integration (Pinheiro and Bates, 1995) to calculate

$$f(y) = \int \frac{\exp(-\nu\mu)(\nu\mu)^y}{y!} f(\nu) d\nu$$

which is equivalent to

$$f(y) = \int \frac{\exp[-\exp(\gamma)\mu^*](\exp(\gamma)\mu^*)^y}{y!} \frac{1}{\sqrt{2\pi\sigma^2}} \exp(\frac{\gamma}{2\sigma^2}) d\gamma$$

where $\mu^* = \exp(\mathbf{x}_j\boldsymbol{\beta} - \frac{\sigma^2}{2})$.

### 2.3. Inverse Gaussian distribution of $\nu$

The inverse Gaussian distribution has been used to handle overdispersion of count data (Sichel, 1971, 1973 and 1974; Houggard, Lee, and Whitmore, 1997). Let $\nu$ have an inverse Gaussian distribution $(1, \frac{1}{\alpha})$,

$$f(\nu|\alpha) = \sqrt{\frac{1}{2\pi\alpha\nu^3}} \exp\{-\frac{(\nu-1)^2}{2\alpha\nu}\}. \tag{3}$$

Then $\mathrm{E}(\nu) = 1$ and $\mathrm{Var}(\nu) = \alpha$.

The marginal probability density function of $y$ has no closed form. Shoukri, et al. (2004) developed estimating equations to obtain estimates of fixed effects. The marginal probability density function of $y$ can be obtained by numerical integration. For simplicity, and to obtain a likelihood, we used numerical integration in this paper to calculate

$$f(y) = \int \frac{\exp(-\nu\mu)(\nu\mu)^y}{y!} \sqrt{\frac{1}{2\pi\alpha\nu^3}} \exp\{-\frac{(\nu-1)^2}{2\alpha\nu}\} d\nu.$$

For the three likelihood methods, their log likelihood values are summed over subjects and maximum likelihood estimates (MLEs) are obtained numerically. We refer to these three likelihood approaches as the GM, LN and IG methods.

### 2.4. Sandwich robust estimator

The robust sandwich variance estimator, which was originally introduced by Huber (1967) and White (1982), has been widely used for analyzing overdispersed count data. A so-called "working covariance matrix" is used in the estimation step (Zeger and Liang, 1986). Then the "working covariance matrix" is combined with its corresponding empirical version in a sandwich estimate form for covariance estimation. The robust sandwich estimator does not make distributional assumptions regarding the unobserved heterogeneity.

### 2.5. Quasilikelihood method

Quasilikelihood is also widely used to account for unobserved heterogeneity in count data analysis (McCullagh and Nelder, 1989). The quasilikelihood function is obtained by dividing a log-likelihood function for Poisson distribution by a dispersion parameter $\phi$. Estimation of $\beta$ and its standard error are computed using the quasilikelihood in place of the log likelihood. Quasilikelihood is avaliable in commercial software (e.g., SAS PROC GENMOD with dscale option). The scale parameter is a fixed parameter which can be estimated from data. The variance of $y$ is then

$$\text{Var}(y) = \phi\mu.$$

$\phi < 1$ models underdispersion and $\phi > 1$ models overdispersion. Breslow (1990) evaluated its performance in analyzing overdispersed Poisson regression.

The moments of the mean $(\lambda)$ and marginal outcome $(y)$ for these three likelihood based methods are shown in Table 1. All of the marginal outcomes have quadratic variance function of the form $Var(y) = E(y) + cE(y)^2$ where $c = Var(\nu)$ is a constant depending on model parameters. We simulated overdispersed count data from the Poisson distribution mixture with three different distributions of $\nu$ but with the same level of overdispersion (same value of c) according to Table 1.

## 3. Simulation studies

The simulations are designed to evaluate these five methods of handling overdispersion of count data under different assumptions.

### 3.1. Simulation algorithm

1. Choose the distribution of $\nu$ from one of the three distributions, the gamma, the lognormal and the inverse Gaussian distributions.

2. Generate the dichotomous independent variable $x_j$ with half the observations having zero for the control group and the other half having 1 for the treatment group, j=1 to 500.

3. Select values for the intercept $(\beta_0)$, the treatment effect $(\beta_1)$, and the parameter for the distribution of $\nu$.

4. Generate independently $\nu_j$, j=1 to 500, from the selected distribution of $\nu$ in step 1.

5. Calculate $\lambda_j = \nu_j \exp(\beta_0 + x_j\beta_1)$ and generate randomly $y_j$ from the Poisson $(\lambda_j)$ distribution.

6. Analyze each dataset with each of the five approaches described in section 2.

Steps 1-6 are repeated 500 times (500 replications) for each combination of parameters.

### 3.2. Simulation parameters

For the fixed effects, we choose 0.693 for $\beta_0$ in all simulations so that the baseline rate is $\lambda = \exp(0.693) = 2$. We consider two values for $\beta_1$, 0.3 and 0.5. This range of $\beta_1$ will allow us to illustrate the difference in empirical power between the five approaches. $\beta_1 = 0$ is also studied for comparing type I error rates.

For the distribution of $\nu$, five values of the distribution parameter were set up to represent different levels of overdispersion. We also use the facts in Table 1 to generate overdispersed count data with the same level of overdispersion but from different distributions of $\nu$ while $\beta_0$ and $\beta_1$ are fixed. For example, count data with $\nu$ having a gamma distribution with $\tau = 0.5$ have the same mean and variance as count data with $\nu$ having a lognormal distribution with $\sigma^2 = 0.405$, and count data with $\nu$ having an inverse Gaussian distribution with $\alpha = 0.5$.

### 3.3. Evaluation measures

These five approaches are compared based on the type I error rates, the bias of estimates of $\beta_1$, standard error of $\hat{\beta}_1$, standard deviation of $\hat{\beta}_1$'s, empirical power, and rates of correct model choice.

1. Type I error rates. The type I error rates are computed as the proportion of p-values less than 0.05 under a null hypothesis of no intervention effect ($\beta_1 = 0$) based on Wald tests.

2. Bias. The bias is calculated as the average of the 500 estimates of $\beta_1$ minus the pre-specified true values. Negative bias represents underestimation of the true value and positive bias represents overestimation of the true values.

3. The mean of standard errors of $\hat{\beta}_1$, standard deviation of $\hat{\beta}_1$'s, and empirical power. The mean of standard errors of $\hat{\beta}_1$ is calculated as the average of the 500 estimates of standard errors of $\hat{\beta}_1$. The standard deviation of $\hat{\beta}_1$'s is the standard deviation of the 500 estimates of $\beta_1$. Empirical power is the proportion of the 500 replications with p-values less than 0.05 under the alternative hypothesis for a given value of $\beta_1$. The estimate of $\beta_1$ and its standard error are factors determining the p-value, the significance of difference between the control and treatment groups, which is generally the main interest of a study.

4. Rates of choosing the true models are calculated as the percentage of those replications with the -2 log likelihood value using the true distribution of $\nu$ less than the -2 log likelihood value using the other two distributions.

## 4. Results

### 4.1. Type I error rate

Type I error rates are shown in Table 2. When the true distribution of $\nu$ is the gamma, the GM, LN, IG methods, and the robust sandwich estimator yield comparable and appropriate type I error rates (around 0.05) while inflated type I errors rates for quasilikelihood increase with increasing value of $\tau$. When the true distribution of $\nu$ is the lognormal or inverse Gaussian, the LN, IG and robust sandwich estimator give similar and appropriate type I error rates (around 0.05) while the GM and quasilikelihood methods have larger type I error rates with higher levels of overdispersion, in some cases well above 0.05.

### 4.2. Bias

Only when $\nu$ has a gamma distribution with $\tau \geq 4$, the LN and IG methods underestimate $\beta_1$ (Table 3). The reason for this is unknown. Under other assumptions, all these five approaches produce ignorable biases. This is expected since unobserved heterogeneity is thought to have an impact on the efficiency of estimation, not on estimate itself.

### 4.3. The mean of standard errors of $\hat{\beta}_1$, standard deviation of $\hat{\beta}_1$'s, and empirical power.

The mean of standard errors of $\hat{\beta}_1$, standard deviation of $\hat{\beta}_1$'s, and empirical power are reported in Tables 4 and 5. The mean of standard errors of $\hat{\beta}_1$ and standard deviation of $\hat{\beta}_1$'s do not change with different values of $\beta_1$. Only $\beta_1 = 0.3$ is reported in Table 4. It is thought that standard deviation of $\hat{\beta}_1$'s is the true representative of the standard errors of $\hat{\beta}_1$. Thus the model with smallest standard deviation of $\hat{\beta}_1$'s and the mean of standard errors of $\hat{\beta}_1$ close to that is considered the best. When the underlying distribution of $\nu$ is gamma and $\tau \geq 4$, the GM and RS methods are considered as appropriate and they gave similar standard errors of $\hat{\beta}_1$ and empirical powers. The LN and IG methods overestimate the standard errors of $\hat{\beta}_1$. Subsequently the LN and IG methods have smaller power which can be up to 0.23 less than the GM method. Although the IG method overestimates the standard error of $\hat{\beta}_1$ less than the LN method, it underestimates $\beta_1$ more than the LN method. It turns out that the LN and IG methods yield a similar level of empirical power. In contrast, the quasilikelihood method underestimates the standard error and gives inflated type I error and empirical power which can be up to 0.21 greater than the GM method.

When the underlying distribution of $\nu$ is lognormal and $\sigma^2 \geq 1.609$, the LN and IG methods are considered appropriate and they gave the similar standard errors and empirical powers. The GM and quasilikelihood methods slightly underestimate the standard errors of $\hat{\beta}_1$ and give empirical powers slightly greater than the LN method. In contrast, the robust sandwich estimator overestimates the standard errors of $\hat{\beta}_1$ and gives empirical powers smaller than the LN method (up to 0.17).

When the underlying distribution of $\nu$ is inverse Gaussian and $\alpha \geq 4$, the LN and IG methods are considered appropriate and they gave the similar standard errors and empirical powers. The robust sandwich estimator overestimates the standard errors of $\hat{\beta}_1$ and gives empirical powers smaller than the IG method (up to 0.15). In contrast, the quasilikelihood method slightly underestimates the standard errors of $\hat{\beta}_1$ and gives empirical powers slightly greater than the IG method (up to 0.12). The GM, LN, and IG methods produce comparable standard errors of $\hat{\beta}_1$ and empirical powers.

### 4.4. Rates of choosing the true models based on -2 log likelihood values

Rates of choosing the true models based on -2 log likelihood values (Table 6) showed that the -2 log likelihood value is a good criterion to determine which model is more appropriate, especially when count data are very overdispersed. From Tables 2, 3 and 4, that is also when $\beta_1$ is underestimated, and when the standard errors of $\hat{\beta}_1$ is under or over-estimated, and when higher type I error rate can arise.

## 5. Discussion

In this simulation study, we demonstrate that 1) there is little bias in any of the methods considered. Only when the true distribution of $\nu$ is gamma and $\text{Var}(\nu) \geq 4.0$, the LN and IG methods underestimate $\beta_1$; 2) different approaches do have impact on the estimated standard errors of $\hat{\beta}_1$, and on type I error rate and empirical power. Specifically, 3) when the count data are mildly overdispersed (e.g., $\text{Var}(\nu) \leq 2.0$), there are virtually no differences in type I error rate, standard error of $\hat{\beta}_1$, and empirical power among the five methods; 4) when the count data are very overdispersed (e.g., $\text{Var}(\nu) \geq 4.0$), none of these five approaches is robust to misspecification of the underlying distribution of $\nu$ as evaluated by type I error rate, the standard error of $\hat{\beta}_1$, and empirical power; 5) -2 log likelihood values provide a good way of choosing the method for dealing with very overdispersed count data (e.g., $\text{Var}(\nu) \geq 4.0$). With our sample size of n=500, over 95 percent of the time, one would choose the true model between the GM method and one of the LN or IG methods. The LN and IG methods are less distinguishable.

Because quasilikelihood has high type I error rate and inflated power, and the robust sandwich estimator gives smaller empirical power for very overdispersed count data, we recommend use of likelihood approaches. Unlike non-likelihood approaches, likelihood approaches allow for statistical tests based on likelihood ratios and for checking model fit and provide basis for power and sample size calculations. There are also practical considerations in choosing among the three Poisson distribution mixtures. The likelihood for the GM method can be computed using the standard closed form expressions for the negative binomial distribution, as in Section 2.1. The likelihood for the LN method (Section 2.2) requires a numerical integration, but since this method can be viewed as a Poisson generalized linear model with log link and a normal random effect in the linear predictor, this calculation is available in some standard softwares (e.g., SAS PROC NLMIXED). The likelihood for the IG method (Section 2.3) also requires a numerical integration which is not available in standard software. Only when a very long and slightly heavy tail exists in count data, the inverse Gaussian has some advantage (Wilmot, 1987). This simulation study raises caution on using non-likelihood method for analyzing very overdispered count data because of likely higher type I error and inappropriate power levels. When likelihood approaches are used, likelihood ratio tests have been used for testing overdispersion. Pearson Statistic and Deviance Statistic are available for assessing overall goodness-of-fit of a model for fitting count data (Cameron and Trivedi, 1998). However, these three likelihood approaches may not be distinguishable with Pearson Statistic and Deviance Statistic. We propose comparing likelihood values to select the appropriate parametric method for analyzing very overdispersed count data. This is equivalent to AIC (Akaike, 1974) because the number of parameters in these three parametric models is the same. Results in Tables 2, 4 and 5 indicate that the standard error of $\hat{\beta}_1$ or empirical power using the incorrect model are a potential problem when count data are very overdispersed, but this approach has an excellent chance of selecting the most appropriate model.

# References

[1] AKAIKE, H. (1974). A new look at the statistical model identification. *IEEE Trans on Automatic Control* **19** 716–722.

[2] BRESLOW, N. (1990). Test of hypotheses in overdispersion regression and other quasilikelihood models. *Journal of the American Statistical Association* **85** 565–571.

[3] CAMERON, A.C., TRIVEDI, P.K. (1998). *Regression Analysis of Count Data.* Cambridge University Press.

[4] CASELLA, G., BERGER, R.L. (1990). *Statistical Inference.* Wadsworth and Brooks.

[5] COX, D.R. (1983). Some remarks on overdispersion. *Biometrika* **70** 269–274.

[6] EFRON, B. (1986). Double exponential families and their use in generalized linear regression. *Journal of the American Statistical Association* **81** 709–721.

[7] ESTABROOKS, P.A., NELSON, C.C., XU, S., KING, D., BAYLISS, E.A., GAGLIO, B., NUTTING, P.A., GLASGOW, R.E. (2005). The frequency and behavioral outcomes of goal choices in the self-management of diabetes. *The Diabetes Educator* **31** 391–400.

[8] FADDY, M.J. (1997). Extended Poisson process modeling and analysis of count data. *Biometrical Journal* **39** 431–440.

[9] HINDE, J. (1982). Compound Poisson regression models. *Lecture Notes in Statistics* **14** 109–121.

[10] HOUGGARD, P., LEE, M.T., WHITMORE, G.A. (1997). Analysis of overdispersed count data by mixtures of Poisson variables and Poisson processes. *Biometrics* **53** 1225-1238.

[11] HUBER, P. J. (1967). *The behavior of maximum likelihood estimation under nonstandard conditions. Proceedings of the Fifth Berkeley Symposium on Mathematical Statistics and Probability 1, LeCam, L. M. and Neyman, J. editors.* UNIVERSITY OF CALIFORNIA PRESS pp. 221-233.

[12] LAWLESS, J.F. (1987). Negative binomial and mixed Poisson regression. *The Canadian Journal of Statistics* **15** 209-225.

[13] LEE, Y., NELDER, J.A. (2000). Two ways of modeling overdispersion in non-normal data. *Journal of the Royal Statistical Society Series C, Applied Statistics* **49** 591–598.

[14] MANTON, K. G., WOODBURY, M. A,, STALLARD, E. (1981). A variance components approach to categorical data models with heterogeneous cell populations: analysis of spatial gradients in lung cancer mortality rates in north Carolina counties. *Biometrics* **37** 259–269.

[15] MARGOLIN, B. H., KAPLAN, N., ZEIGER, E. (1981). Statistical analysis of the Ames Salmonella Microsome Test. *Proceedings of the National Academy of Sciences* **76** 3779–3783.

[16] BILLINGSLEY, P. (1999). *Convergence of Probability Measures*, 2nd ed. Wiley, New York.

[17] MCCULLAGH, P., NELDER, J.A. (1989). *Generalized Linear Models.* London and Boca Raton, Florida: Chapman & Hall/CRC.

[18] NEAS, L.M., SCHWARTZ, J., DOCKERY, D. (1999). A case-crossover analysis of air pollution and mortality in Philadelphia. *Environmental Health Perspectives* **107** 629–631.

[19] PAUL, S.R., PLACKETT, R.L. (1978). Inference sensitivity for Poisson mixtures. *Biometrika* **65** 591-602.

[20] PINHEIRO, J.C., BATES, D.M. (1995). Approximations to the loglikelihood function in the nonlinear mixed-effects model. *Journal of Computational and Graphical Statistics* **4** 12–35.

[21] RIDOUT, M.S., BESBEAS, P. (2004). An empirical model for underdispersed count data. *Statistical Modeling* **4** 77-89.

[22] SHOUKRI, M.M., ASYALI, M.H., VANDORP, R., KELTON, D. (2004). The Poisson inverse Gaussian regression model in the analysis of clustered counts data. *Journal of data Science* **2** 17–32.

[23] SICHEL, H.S. (1971). *On a family of discrete distributions particularly suited to represent long-tailed frequency data.* In: Laubscher,N.F. ed., *Proceedings of the Third Symposium on Mathematical Statistics. Pretoria CSIR.*

[24] SICHEL, H.S. (1973). The density and size distribution of diamonds. *Bulletin of the International Statistical Institute* **45(2)** 420–427.

[25] SICHEL, H.S. (1974). On a distribution representing sentence-length in written prose. *Journal of the Royal Statistical Society, Series A* **137** 25–34.

[26] SILLS, M.R., SHETTERLY, S., XU, S., MAGID, D., KEMPE, A. (2007). The association between parental depression and children's healthcare utilization. *Pediatrics* **119** 829–836.

[27] STIRATELLI, R., LAIRD, N., WARE, J.H. (1984). Random-effects models for serial observations with binary response. *Biometrics* **40** 961–971.

[28] THALL, P.F. VAIL, S.C. (1990). Some covariance models for longitudinal count data with overdispersion. *Biometrics* **46** 657–671.

[29] WHITE, H. (1982). Maximum likelihood estimation of misspecifed models. *Econometrica* **50** 1–25.

[30] WILMOT, G.E. (1987). The Poisson-Inverse Gaussian distribution as an Alternative to the negative binomial. *Scandinavian Actuarial Journal* **1** 13–127.

[31] ZEGER, S.L. LIANG, K.Y. (1986). Longitudinal data analysis for discrete and continuous outcomes. *Biometrics* **42** 121–130.

TABLE 1
*Moments of the mean ($\lambda$) and marginal outcome ($y$)*

| moments | distribution of $\nu$ | | |
|---|---|---|---|
| mean or variance | gamma | lognormal | inverse Gaussian |
| $E(\lambda)$ | $\mu$ | $\mu$ | $\mu$ |
| $V(\lambda)$ | $\tau\mu^2$ | $(\exp(\sigma^2)-1)\mu^2$ | $\alpha\mu^2$ |
| $E(y)$ | $\mu$ | $\mu$ | $\mu$ |
| $V(y)$ | $\mu+\tau\mu^2$ | $\mu+(\exp(\sigma^2)-1)\mu^2$ | $\mu+\alpha\mu^2$ |

TABLE 2
*Type I error rates by different distributions of $\nu$ and different levels of overdispersion based on 500 replications*

| true dist'n of $\nu$ | parm | likelihood | | | non-likelihood | |
|---|---|---|---|---|---|---|
| | | GM | LN | IG | RS | QL |
| GM | $\tau$=0.5 | 0.046 | 0.032 | 0.054 | 0.054 | 0.052 |
| | $\tau$=1.0 | 0.056 | 0.052 | 0.024 | 0.058 | 0.052 |
| | $\tau$=2.0 | 0.054 | 0.042 | 0.042 | 0.048 | 0.086 |
| | $\tau$=4.0 | 0.042 | 0.050 | 0.056 | 0.050 | 0.128 |
| | $\tau$=6.0 | 0.034 | 0.042 | 0.046 | 0.046 | 0.156 |
| LN | $\sigma^2 = 0.405$ | 0.050 | 0.042 | 0.050 | 0.042 | 0.046 |
| | $\sigma^2 = 0.693$ | 0.072 | 0.056 | 0.054 | 0.054 | 0.076 |
| | $\sigma^2 = 1.098$ | 0.088 | 0.044 | 0.042 | 0.056 | 0.098 |
| | $\sigma^2 = 1.609$ | 0.130 | 0.054 | 0.062 | 0.050 | 0.158 |
| | $\sigma^2 = 1.946$ | 0.130 | 0.052 | 0.060 | 0.058 | 0.186 |
| IG | $\alpha$=0.5 | 0.052 | 0.042 | 0.038 | 0.058 | 0.060 |
| | $\alpha$=1.0 | 0.064 | 0.052 | 0.044 | 0.052 | 0.074 |
| | $\alpha$=2.0 | 0.104 | 0.062 | 0.058 | 0.066 | 0.110 |
| | $\alpha$=4.0 | 0.124 | 0.058 | 0.034 | 0.056 | 0.168 |
| | $\alpha$=6.0 | 0.148 | 0.064 | 0.066 | 0.064 | 0.196 |

RS=robust sandwich estimator; QL=quasilikelihood method.

TABLE 3
*Mean bias by different distributions of $\nu$ and different levels of overdispersion based on 500 replications*

| true | | | likelihood | | | non-likelihood | |
|------|------|--------|--------|--------|--------|--------|--------|
| dist'n of $\nu$ | parm | $\beta_1$ | GM | LN | IG | RS | QL |
| GM | $\tau$=0.5 | 0.3 | 0.005 | 0.005 | 0.004 | 0.005 | 0.005 |
| | | 0.5 | 0.006 | 0.004 | 0.003 | 0.006 | 0.006 |
| | $\tau$=1.0 | 0.3 | 0.000 | 0.000 | -0.004 | 0.000 | 0.000 |
| | | 0.5 | 0.001 | -0.001 | -0.007 | 0.001 | 0.001 |
| | $\tau$=2.0 | 0.3 | 0.011 | 0.006 | 0.007 | 0.011 | 0.011 |
| | | 0.5 | 0.010 | 0.005 | 0.007 | 0.010 | 0.010 |
| | $\tau$=4.0 | 0.3 | -0.013 | -0.041 | -0.062 | -0.013 | -0.013 |
| | | 0.5 | 0.005 | -0.030 | -0.077 | 0.005 | 0.005 |
| | $\tau$=6.0 | 0.3 | -0.015 | -0.037 | -0.074 | -0.015 | -0.015 |
| | | 0.5 | 0.004 | -0.034 | -0.105 | 0.004 | 0.004 |
| LN | $\sigma^2 = 0.405$ | 0.3 | 0.001 | 0.001 | 0.001 | 0.001 | 0.001 |
| | | 0.5 | -0.002 | -0.001 | -0.002 | -0.002 | -0.002 |
| | $\sigma^2 = 0.693$ | 0.3 | 0.002 | 0.001 | 0.002 | 0.002 | 0.002 |
| | | 0.5 | 0.001 | 0.000 | 0.001 | 0.001 | 0.001 |
| | $\sigma^2 = 1.098$ | 0.3 | 0.001 | 0.002 | 0.006 | 0.001 | 0.001 |
| | | 0.5 | 0.002 | 0.007 | 0.006 | 0.002 | 0.002 |
| | $\sigma^2 = 1.609$ | 0.3 | 0.003 | 0.009 | 0.006 | 0.004 | 0.004 |
| | | 0.5 | -0.006 | -0.003 | 0.002 | -0.004 | -0.004 |
| | $\sigma^2 = 1.946$ | 0.3 | -0.005 | 0.003 | 0.002 | -0.003 | -0.003 |
| | | 0.5 | -0.006 | 0.006 | 0.002 | 0.006 | 0.006 |
| IG | $\alpha$=0.5 | 0.3 | 0.002 | 0.000 | 0.000 | 0.002 | 0.002 |
| | | 0.5 | 0.001 | 0.000 | 0.000 | 0.001 | 0.001 |
| | $\alpha$=1.0 | 0.3 | -0.003 | -0.004 | -0.004 | -0.003 | -0.003 |
| | | 0.5 | 0.001 | 0.001 | 0.006 | 0.001 | 0.001 |
| | $\alpha$=2.0 | 0.3 | 0.011 | 0.006 | 0.000 | 0.011 | 0.011 |
| | | 0.5 | 0.010 | 0.005 | 0.007 | 0.010 | 0.010 |
| | $\alpha$=4.0 | 0.3 | 0.011 | 0.002 | 0.001 | 0.011 | 0.011 |
| | | 0.5 | 0.010 | -0.002 | 0.000 | 0.010 | 0.010 |
| | $\alpha$=6.0 | 0.3 | 0.013 | 0.009 | 0.012 | 0.014 | 0.014 |
| | | 0.5 | 0.010 | 0.006 | 0.008 | 0.013 | 0.013 |

RS=robust sandwich estimator; QL=quasilikelihood method.

TABLE 4

*The mean of standard errors of $\hat{\beta}_1$ (standard deviation of $\hat{\beta}_1$'s) by different distributions of $\nu$ and different levels of overdispersion based on 500 replications for $\beta_1 = 0.3$*

| true dist'n of $\nu$ | parm | likelihood | | | non-likelihood | |
|---|---|---|---|---|---|---|
| | | GM | LN | IG | RS | QL |
| GM | $\tau = 0.5$ | 0.086 | 0.088 | 0.088 | 0.086 | 0.087 |
| | | (0.088) | (0.090) | (0.089) | (0.088) | (0.088) |
| | $\tau = 1.0$ | 0.107 | 0.112 | 0.111 | 0.107 | 0.103 |
| | | (0.106) | (0.112) | (0.111) | (0.106) | (0.106) |
| | $\tau = 2.0$ | 0.139 | 0.153 | 0.147 | 0.138 | 0.122 |
| | | (0.141) | (0.152) | (0.149) | (0.141) | (0.141) |
| | $\tau = 4.0$ | 0.188 | 0.219 | 0.201 | 0.187 | 0.147 |
| | | (0.181) | (0.209) | (0.195) | (0.181) | (0.181) |
| | $\tau = 6.0$ | 0.226 | 0.274 | 0.238 | 0.223 | 0.161 |
| | | (0.219) | (0.269) | (0.238) | (0.219) | (0.219) |
| LN | $\sigma^2 = 0.405$ | 0.084 | 0.085 | 0.085 | 0.086 | 0.085 |
| | | (0.087) | (0.086) | (0.085) | (0.087) | (0.087) |
| | $\sigma^2 = 0.693$ | 0.098 | 0.101 | 0.101 | 0.105 | 0.097 |
| | | (0.106) | (0.100) | (0.102) | (0.108) | (0.108) |
| | $\sigma^2 = 1.098$ | 0.115 | 0.121 | 0.120 | 0.134 | 0.112 |
| | | (0.136) | (0.120) | (0.121) | (0.136) | (0.136) |
| | $\sigma^2 = 1.609$ | 0.135 | 0.143 | 0.139 | 0.172 | 0.128 |
| | | (0.179) | (0.145) | (0.149) | (0.182) | (0.182) |
| | $\sigma^2 = 1.946$ | 0.148 | 0.157 | 0.157 | 0.199 | 0.138 |
| | | (0.206) | (0.163) | (0.164) | (0.213) | (0.213) |
| IG | $\alpha = 0.5$ | 0.084 | 0.085 | 0.085 | 0.086 | 0.085 |
| | | (0.090) | (0.086) | (0.086) | (0.090) | (0.090) |
| | $\alpha = 1.0$ | 0.100 | 0.103 | 0.103 | 0.107 | 0.099 |
| | | (0.109) | (0.107) | (0.103) | (0.109) | (0.109) |
| | $\alpha = 2.0$ | 0.121 | 0.128 | 0.125 | 0.138 | 0.116 |
| | | (0.142) | (0.131) | (0.131) | (0.142) | (0.142) |
| | $\alpha = 4.0$ | 0.149 | 0.159 | 0.159 | 0.186 | 0.137 |
| | | (0.188) | (0.164) | (0.156) | (0.188) | (0.188) |
| | $\alpha = 6.0$ | 0.168 | 0.180 | 0.179 | 0.220 | 0.149 |
| | | (0.229) | (0.185) | (0.182) | (0.230) | (0.230) |

RS=robust sandwich estimator; QL=quasilikelihood method.

TABLE 5

*Empirical power by different distributions of $\nu$ and different levels of overdispersion based on 500 replications*

| true dist'n of $\nu$ | parm | $\beta_1$ | likelihood | | | non-likelihood | |
|---|---|---|---|---|---|---|---|
| | | | GM | LN | IG | RS | QL |
| GM | $\tau=0.5$ | 0.3 | 0.94 | 0.93 | 0.93 | 0.94 | 0.94 |
| | | 0.5 | 1.00 | 1.00 | 1.00 | 1.00 | 1.00 |
| | $\tau=1.0$ | 0.3 | 0.81 | 0.77 | 0.77 | 0.82 | 0.82 |
| | | 0.5 | 1.00 | 1.00 | 1.00 | 1.00 | 1.00 |
| | $\tau=2.0$ | 0.3 | 0.59 | 0.49 | 0.49 | 0.59 | 0.68 |
| | | 0.5 | 0.95 | 0.88 | 0.87 | 0.95 | 0.97 |
| | $\tau=4.0$ | 0.3 | 0.31 | 0.22 | 0.21 | 0.33 | 0.51 |
| | | 0.5 | 0.77 | 0.58 | 0.57 | 0.76 | 0.86 |
| | $\tau=6.0$ | 0.3 | 0.24 | 0.14 | 0.14 | 0.24 | 0.43 |
| | | 0.5 | 0.61 | 0.38 | 0.38 | 0.61 | 0.82 |
| LN | $\sigma^2=0.405$ | 0.3 | 0.95 | 0.95 | 0.94 | 0.94 | 0.94 |
| | | 0.5 | 1.00 | 1.00 | 1.00 | 1.00 | 1.00 |
| | $\sigma^2=0.693$ | 0.3 | 0.85 | 0.86 | 0.85 | 0.78 | 0.85 |
| | | 0.5 | 1.00 | 1.00 | 1.00 | 0.99 | 1.00 |
| | $\sigma^2=1.098$ | 0.3 | 0.73 | 0.71 | 0.70 | 0.62 | 0.75 |
| | | 0.5 | 0.99 | 0.99 | 0.99 | 0.96 | 0.99 |
| | $\sigma^2=1.609$ | 0.3 | 0.59 | 0.55 | 0.56 | 0.45 | 0.64 |
| | | 0.5 | 0.95 | 0.95 | 0.93 | 0.83 | 0.93 |
| | $\sigma^2=1.946$ | 0.3 | 0.52 | 0.48 | 0.48 | 0.35 | 0.56 |
| | | 0.5 | 0.85 | 0.89 | 0.88 | 0.71 | 0.86 |
| IG | $\alpha=0.5$ | 0.3 | 0.94 | 0.94 | 0.93 | 0.94 | 0.94 |
| | | 0.5 | 1.00 | 1.00 | 1.00 | 1.00 | 1.00 |
| | $\alpha=1.0$ | 0.3 | 0.82 | 0.81 | 0.80 | 0.79 | 0.83 |
| | | 0.5 | 1.00 | 1.00 | 1.00 | 0.99 | 1.00 |
| | $\alpha=2.0$ | 0.3 | 0.68 | 0.65 | 0.66 | 0.60 | 0.71 |
| | | 0.5 | 0.97 | 0.97 | 0.97 | 0.95 | 0.98 |
| | $\alpha=4.0$ | 0.3 | 0.54 | 0.48 | 0.50 | 0.40 | 0.60 |
| | | 0.5 | 0.89 | 0.89 | 0.90 | 0.78 | 0.89 |
| | $\alpha=6.0$ | 0.3 | 0.48 | 0.42 | 0.44 | 0.32 | 0.56 |
| | | 0.5 | 0.79 | 0.79 | 0.80 | 0.64 | 0.83 |

RS=robust sandwich estimator; QL=quasilikelihood method.

TABLE 6
*Rates of choosing the true models based on -2 log likelihood values and 500 replications*

| true | | | assumed dist'n of $\nu$ | |
|---|---|---|---|---|
| distribution of $\nu$ | parameter | $\beta_1$ | GM versus LN | GM versus IG |
| GM | $\tau$=0.5 | 0.3 | 0.85 | 0.83 |
| | | 0.5 | 0.87 | 0.84 |
| | $\tau$=1.0 | 0.3 | 0.96 | 0.95 |
| | | 0.5 | 0.98 | 0.96 |
| | $\tau$=2.0 | 0.3 | 0.99 | 0.99 |
| | | 0.5 | 1.00 | 1.00 |
| | $\tau$=4.0 | 0.3 | 1.00 | 1.00 |
| | | 0.5 | 1.00 | 0.99 |
| | $\tau$=6.0 | 0.3 | 1.00 | 0.99 |
| | | 0.5 | 1.00 | 0.99 |
| distribution of $\nu$ | parameter | $\beta_1$ | LN versus GM | LN versus IG |
| LN | $\sigma^2 = 0.405$ | 0.3 | 0.78 | 0.44 |
| | | 0.5 | 0.77 | 0.43 |
| | $\sigma^2 = 0.693$ | 0.3 | 0.88 | 0.52 |
| | | 0.5 | 0.88 | 0.53 |
| | $\sigma^2 = 1.098$ | 0.3 | 0.95 | 0.59 |
| | | 0.5 | 0.97 | 0.60 |
| | $\sigma^2 = 1.609$ | 0.3 | 0.95 | 0.59 |
| | | 0.5 | 0.96 | 0.60 |
| | $\sigma^2 = 1.946$ | 0.3 | 0.95 | 0.59 |
| | | 0.5 | 0.94 | 0.64 |
| distribution of $\nu$ | parameter | $\beta_1$ | IG versus GM | IG versus LN |
| IG | $\alpha$=0.5 | 0.3 | 0.77 | 0.68 |
| | | 0.5 | 0.79 | 0.67 |
| | $\alpha$=1.0 | 0.3 | 0.91 | 0.72 |
| | | 0.5 | 0.92 | 0.75 |
| | $\alpha$=2.0 | 0.3 | 0.98 | 0.81 |
| | | 0.5 | 0.97 | 0.81 |
| | $\alpha$=4.0 | 0.3 | 0.99 | 0.89 |
| | | 0.5 | 0.99 | 0.87 |
| | $\alpha$=6.0 | 0.3 | 0.99 | 0.95 |
| | | 0.5 | 0.99 | 0.94 |

RS=robust sandwich estimator; QL=quasilikelihood method.